\documentstyle[epsf]{article}
\makeatletter

\@addtoreset{equation}{section}
\makeatother
\begin{document}
\baselineskip 0.8cm

\newcommand{\gsim}{ \mathop{}_{\textstyle \sim}^{\textstyle >} }
\newcommand{\lsim}{ \mathop{}_{\textstyle \sim}^{\textstyle <} }
\newcommand{\vev}[1]{ \left\langle {#1} \right\rangle }
\newcommand{\EV}{ {\rm eV} }
\newcommand{\KEV}{ {\rm keV} }
\newcommand{\MEV}{ {\rm MeV} }
\newcommand{\GEV}{ {\rm GeV} }
\newcommand{\TEV}{ {\rm TeV} }
\newcommand{\DS}{\displaystyle}
\def\tr{\mathop{\rm tr}\nolimits}
\def\Tr{\mathop{\rm Tr}\nolimits}
\def\Re{\mathop{\rm Re}\nolimits}
\def\Im{\mathop{\rm Im}\nolimits}
\def\simgt{\mathop{>}\limits_{\displaystyle{\sim}}}
\def\simlt{\mathop{<}\limits_{\displaystyle{\sim}}}
\setcounter{footnote}{0}

\begin{flushright}
KEK-TH-944\\
\end{flushright}

\vskip 1.5cm
\begin{center}
{\large \bf
Probing complex phase\\
of the photon-photon-Higgs vertex\\
at photon colliders
\footnote{based on the work in collaboration with 
\uppercase{K}.~\uppercase{H}agiwara.\uppercase{T}alk presented
at {\it \uppercase{SUSY} 2003:
\uppercase{S}upersymmetry in the \uppercase{D}esert}\/, 
held at the \uppercase{U}niversity of \uppercase{A}rizona,
\uppercase{T}ucson, \uppercase{AZ}, \uppercase{J}une 5-10, 2003.
\uppercase{T}o appear in the \uppercase{P}roceedings.}}
\vskip 0.8cm

 Eri Asakawa

{\it Theory Group, KEK,Tsukuba, Ibaraki 305-0801, Japan\\ 
E-mail: eri@post.kek.jp}
\vskip 0.8cm

\abstract{
We study the effect of heavy neutral Higgs bosons on the 
$t\overline{t}$ production process at photon linear colliders.
The interference patterns between the resonant Higgs production
amplitudes and the continuum QED amplitudes are examined.
The patterns tell us not only the CP nature of the Higgs bosons
but also the phase of the $\gamma\gamma$-Higgs vertex which
gives new information about the Higgs couplings to new charged
particles. }

\end{center}
\vskip 0.8cm

\renewcommand{\thefootnote}{\arabic{footnote}}
\setcounter{footnote}{0}

\section{Introduction}

Although the standard model is consistent with the current experimental
data, 
new physics will be indispensable
if we consider the hierarchy
between the electroweak scale and the Planck scale,
a failure of the gauge coupling unification {\it etc}.
as serious problems.
The new physics may contribute significantly to the Higgs
production at photon colliders, especially through
new charged particles,
because neutral Higgs bosons are produced 
as $s$-channel resonances via loops of charged massive particles.
Then, the effects appear in the $\gamma\gamma$-Higgs
vertex which generally has complex phase even without CP violation.

In this report, we study the interference patterns of the resonant 
and the continuum amplitudes
for the $\gamma\gamma \rightarrow
t \overline{t}$ process by using the circularly 
polarized colliding photons~\cite{ah}. 
It will be shown that
these interference effects allow us to observe the complex phase of 
the $\gamma\gamma$-Higgs vertices, as has been shown
in $WW$ and $ZZ$ production processes~\cite{NZK}.

\section{Helicity amplitudes for the process $\gamma \gamma 
\rightarrow t \overline{t}$}

When the $\gamma\gamma$ collision energy reaches around the
mass of a spinless boson $\phi$ ($\phi=H$ or $A$ where $H$ and 
$A$ are the CP-even and CP-odd Higgs bosons respectively.),
the helicity amplitudes for the process $\gamma_{\lambda_1}
\gamma_{\lambda_2} \rightarrow
t_\sigma \overline{t}_{\overline{\sigma}}$ can be expressed as
\begin{eqnarray}
\label{ampeq}
{\it M}_{\lambda_1 \lambda_2}^{\sigma \overline{\sigma}} =
\left[ {\it M}_\phi \right]
_{\lambda_1 \lambda_2}^{\sigma \overline{\sigma}} +
\left[ {\it M}_t \right]
_{\lambda_1 \lambda_2}^{\sigma \overline{\sigma}} ,
\end{eqnarray}
where the first term ${\it M}_\phi$ stands for the $s$-channel
$\phi$-exchange amplitudes and the latter term ${\it M}_t$
stands for the $t$- and $u$-channel top-quark-exchange amplitudes.
The resonant helicity amplitudes are calculated by using
the lowest-dimensional effective
Lagrangian of the form
\begin{eqnarray}
{\it L}_{\phi \gamma \gamma} &=& \frac{1}{m_\phi}
\left( b_\gamma^H A_{\mu\nu} A^{\mu\nu} +
b_{\gamma}^A \widetilde{A}_{\mu\nu} A^{\mu\nu} \right) \phi.
\end{eqnarray}

By considering the decay angular distribution of $t\overline{t}$
pairs, we can derive 
the convoluted
four observables, $\Sigma_1$ to $\Sigma_4$,
\begin{eqnarray}
\Sigma_i(\sqrt{s}_{\gamma\gamma})=
\int d\sqrt{s}_{\gamma\gamma} \sum_{\lambda_1,~\lambda_2}
\left( \frac{1}{{\it L}_{0.8}} 
\frac{d{\it L}^{\lambda_1\lambda_2}}
{d\sqrt{s}_{\gamma\gamma}} \right) \left( \frac{3\beta}
{32\pi s_{\gamma\gamma}}
\int S^i_{\lambda_1 \lambda_2} (\Theta, \sqrt{s}_{\gamma\gamma})d\cos\Theta
\right),
\\ \nonumber
~~~~~~~~~~~~~~~~~~~~~~~~~~~~~~~~~~~~~~~~~~~~~~~~~{\rm for}~~i=1-4,
\end{eqnarray}
where the functions $S_{\lambda_1 \lambda_2}^i$ contain
all the information about the $\gamma\gamma \rightarrow 
t\overline{t}$ helicity amplitudes:
\begin{eqnarray}
\label{S1toS4}
&&S^1_{\lambda_1 \lambda_2} = 
\left| {\it M}_{\lambda_1\lambda_2}^{RR} \right|^2,~~~~~~
S^2_{\lambda_1 \lambda_2} = \left|{\it M}_{\lambda_1\lambda_2}^{LL}\right|^2,
\\ \nonumber
&&S^3_{\lambda_1 \lambda_2} =
2\Re\left[{\it M}_{\lambda_1\lambda_2}^{RR} \left({\it M}
_{\lambda_1\lambda_2}^{LL}\right)^*\right],~~~
S^4_{\lambda_1 \lambda_2} =
2\Im\left[{\it M}_{\lambda_1\lambda_2}^{RR} \left({\it M}
_{\lambda_1\lambda_2}^{LL}\right)^*\right].
\end{eqnarray}
$\Theta$ is the polar angle of the $t$ momentum in the
$\gamma\gamma$ CM frame, and
the normalized luminosity distribution for each
$\gamma\gamma$ helicity combination is expressed by 
$(1/{\it L}_{0.8})
d{\it L}^{\lambda_1 \lambda_2}/d \sqrt{s}_{\gamma\gamma}$.
In this report, the luminosity distribution is derived
by assuming $\sqrt{s}_{ee}=500$ GeV, $x=4.8$,
$P_l=-1.0$ and $P_e=0.9$~\cite{ah}.

\section{Effects of the 
$\gamma\gamma\phi$ phase on the observables}

We study the $\arg(b_\gamma^\phi)$
dependence of the four observables defined in the previous
section. We first re-parameterize the $J_z=0$ amplitudes of
eq.~(\ref{ampeq}) as follows:
\begin{eqnarray}
{\it M}_{\lambda \lambda}^{\sigma \sigma}=
\left[{\it M}_t \right]_{\lambda \lambda}^{\sigma \sigma}
+ \left( \frac{\sqrt{s}_{\gamma\gamma}}{m_\phi} \right)^3
r_\phi \cdot i \left[ 1+{\rm exp}\left(
2i \tan^{-1} \frac{s_{\gamma\gamma}^2-m_\phi^2}{m_\phi \Gamma_\phi} \right)
\right],
\end{eqnarray}
where
$r_\phi \propto b_\gamma^\phi $.
In this expression, the phase of the Breit-Wigner resonance
amplitude is shifted by the phase of the $r_\phi$ factor
which is essentially the phase of the $\gamma\gamma\phi$ vertex
factor $b_\gamma^\phi$ if we neglect the phase in the 
$t \overline{t} \phi$ vertex.
We note here that 
the $\arg(b_\gamma^\phi)$ depend significantly
on the model parameters. As an example, we show in 
Table~\ref{bpara} the MSSM prediction for the real and imaginary 
parts of $b_\gamma^\phi$.
Here, we calculate the $A$ and $H$ masses and couplings for the 
MSSM parameters; $m_A=400$ GeV, $\tan\beta=3$,
$m_{\widetilde{f}}=1$ TeV, $M_2=500$ GeV, $\mu=-500$ GeV. 
Because the imaginary part
of the $\phi \rightarrow \gamma \gamma$ amplitude is a sum of
the contribution from the $\phi$ decay modes into charged particles
whereas the real part receives contribution from all the charged 
particles, we expect that $\arg(b_\gamma^\phi)$ is a good probe
of heavy charged particles.
\begin{table}[ph]
\begin{center}
\caption{An example of the values $b_\gamma^\phi$.
\label{bpara}}
{\footnotesize
\begin{tabular}{|c|c|c|}
\hline
& $b_\gamma^A\times10^4$ & $b_\gamma^H\times10^4$ \\
\hline
 total & $14+12i$  & $11+1.3i$ \\
\hline\hline
 $t$ & $15+12i$ & $12+3.3i$ \\ 
\hline
 $b$ & $-0.19+0.15i$ & $0.18-0.15i$ \\
\hline
 $W$ & 0.0 & $-1.0-1.7i$ \\
\hline
$\widetilde{\chi}_1^-$ & $-1.1$ & $-1.2$ \\
\hline
$\widetilde{\chi}_2^-$ & $0.51$ & $1.0$ \\
\hline 
\end{tabular}}
\end{center}
\vspace*{-13pt}
\end{table}

\begin{figure}[ht]
\centerline{\epsfxsize=5cm\epsfysize=5cm\epsfbox{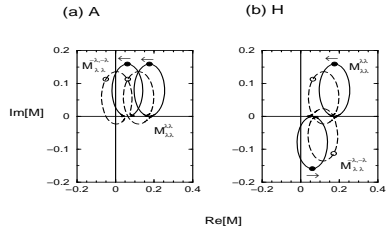}}
\caption{ The $s_{\gamma\gamma}$-dependence of the $\gamma\gamma
\rightarrow t\overline{t}$
amplitudes ${\it M}_{\lambda\lambda}^{\sigma\sigma}$
at $\Theta=0^o$. 
The amplitudes with $A$ production are shown
in the left figure, whereas those for $H$ production
are shown in the right. 
The cases of $arg(b_\gamma^\phi)=0$ and
$\pi/4$ are denoted by
the solid and dashed circles, respectively. 
The small arrows indicate
the direction of increasing $s_{\gamma\gamma}$ and
the solid and open small circles on the trajectories show the
$s_{\gamma\gamma}=m_\phi^2$ points. As $s_{\gamma\gamma}$ grows
the amplitudes make counterclockwise trajectories, and
the magnitude of the resonance amplitude hits its 
maximum at $s_{\gamma\gamma}=m_\phi^2$.}\label{argant}
\end{figure}

Fig.~\ref{argant}
shows plots of the amplitudes ${\it M}_{\lambda \lambda}
^{\sigma \sigma}$ on the complex plane where the scattering angle
$\Theta$ is fixed to be zero as a sample.
\begin{figure}[ht]
\begin{center}
\epsfxsize=4cm
\epsfysize=5cm
\epsffile{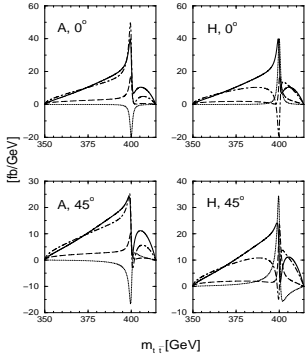}
\caption{The observables $\Sigma_1$ to $\Sigma_4$ 
with no smearing by detector resolution.
The solid, dashed, dot-dashed and dotted curves are
$\Sigma_1$, $\Sigma_2, \Sigma_3$ and $\Sigma_4$, respectively.
The observables with $A$ production are in the left
(right) figures whereas those with the
$H$ production are shown in the right. 
The upper and lower figures 
show the case of
$arg(r_\phi)=0$  and $\pi/4$, respectively. }\label{phase}
\end{center}
\end{figure}
When we compare the $\arg(b_\gamma^A)=0$ amplitudes
(solid circles) and the $\arg(b_\gamma^A)=\pi/4$ amplitudes
(dashed circles), we notice that the magnitudes of all the
amplitudes are reduced for $\arg(b_\gamma^A)>0$ because
the imaginary parts of the resonant amplitudes are positive
for $\arg(b_\gamma^A)=0$. It is notable that at $s_{\gamma\gamma}=m_A^2$
(solid and open circles along the trajectries), the real part of
the ${\it M}_{\lambda\lambda}^{-\lambda,-\lambda}$ amplitudes 
become negative when $\arg(b_\gamma^A)=\pi/4$. In case of
the $\phi=H$ amplitudes shown in Fig.~\ref{argant}(b),
the most notable feature is that the magnitude of the
${\it M}_{\lambda\lambda}^{-\lambda,-\lambda}$ amplitudes
increases for $\arg(b_\gamma^A)>0$ because the sign of the 
imaginary part of the $H$ resonant amplitude is negative
for these amplitudes. On the other hand, the magnitudes of
the ${\it M}_{\lambda\lambda}^{\lambda\lambda}$ amplitudes
decreases for $\arg(b_\gamma^A)>0$ as in the case for the $A$
production amplitudes. 

We show in Fig.~{\ref{phase}} the four observables
$\Sigma_1$ to $\Sigma_4$ for the $A$ production in the left,
and for the $H$ production in the right-hand side.
We find that 
the $\arg(b_\gamma^\phi)$ dependence of
the four observables are significant enough that the phase
of the $\gamma\gamma\phi$ vertex function may be measured
experimentally by a careful study of all the observables. 



\begin{thebibliography}{0}

\bibitem{ah}  E.~Asakawa and K.~Hagiwara, 
{\it Eur. Phys. J.}{\bf C31}, 351 (2003).

\bibitem{NZK}  P.~Niezurawski, A.F.~Zarnecki and M.~Krawczyk, 
{\it JHEP}{\bf 0211}, 034 (2002).

\end{thebibliography}
\end{document}